\documentstyle[spie]{article}

\title{In-orbit performance of the XMM-Newton X-ray telescopes: images and 
spectra}

\author{B. Aschenbach
\skiplinehalf
Max-Planck-Institut f\"ur extraterrestrische Physik, \\
Postfach 1312, 85741 Garching, Germany
}
\authorinfo{\\B.A.:
Email: bra@mpe.mpg.de}
  \begin{document}
  \maketitle
\begin{abstract}
The XMM-Newton X-ray Observatory was launched by an ARIANE V from Kourou,
French Guiana, on 10 December 1999. First light was received by one of the 
three X-ray telescopes on 19 January 2000. Since then an extensive program, 
comprising commissioning, calibration and performance verification of the 
scientific payload, has been carried out, followed by regular scientific 
Guaranteed Time Observations which are interleaved with observations 
drawn from 
 the Guest Observer (AO-1) program. I present the in-orbit
performance of the three X-ray telescopes and demonstrate the excellent 
agreement with the ground calibration measurements. This includes the imaging 
characteristics both on-axis and off-axis, the effective collecting area and 
vignetting making use of in-orbit images and spectra. 
The scientific impact of a few of these observations is highlighted as
well.
\end{abstract}

\keywords{XMM-Newton, X-ray telescopes, X-ray optics testing, X-ray 
straylight}
\section{INTRODUCTION}

On December 10 1999 the XMM spacecraft was placed in a 48h Earth orbit 
by the first commercial ARIANE V launcher. 
XMM, or X-ray Multi-Mirror Mission, re-named XMM-Newton after its 
successful launch, is the second cornerstone of the 
Horizon 2000 science program of the European Space Agency ESA. 
XMM-Newton was designed as a  facility class X-ray astronomy observatory to 
study 
cosmic X-ray sources spectroscopically with the highest collecting area 
feasible in the 0.1 - 10 keV band. 
Over the full band moderate spectral resolving power between 1 and 60 
is required and medium resolving power of $\ge$ 250 was aimed for 
in the 0.1 - 3 keV band. The high throughput is primarily achieved by a set 
of 3 imaging, highly nested Wolter type I telescopes. The imaging performance 
of each of the grazing incidence mirror systems was originally set to 
 be better than 
30 arcsec for the half energy width (HEW) of the point spread function 
(PSF) with a
 goal of 10 arcsec. 
In a first attempt the scientific requirements have been put together at a 
workshop 
held at Lyngby, Denmark, in June 1985$\sp 1$. A detailed analysis of the 
types
of telescopes with which these requirements could be achieved was carried out 
by the ESA established Telescope Working Group, which also arrived at a 
first telescope design$\sp{2,3,4}$. 
Over ten years in time the design was put into reality, the telescopes 
were built in industry including Carl Zeiss, Oberkochen in Germany, 
Kayser-Threde, Munich in Germany and Media Lario, Bosisio Parini in Italy. 
The development and building of the mirror shells up to the flight mirror 
modules was accompanied by X-ray tests 
in the 130 m long X-ray test facility PANTER of the Max-Planck-Institut 
f\"ur extraterrestrische Physik. The qualification of the mirror 
modules was performed in EUV light at the Focal-X facility of the 
Centre Spatial de Li\`ege in Belgium. 
A more detailed overview of the mission, the spacecraft and mission operations 
can be found in reference 5. 
Some analysis and summary 
of the ground calibration of the X-ray telescopes prior to launch is 
provided by Gondoin et al.$\sp{6,7}$. 

One of the X-ray telescopes is equipped with a focal plane imaging camera 
of  pn semiconductor CCD technology$\sp{8, 9}$ 
(EPIC-pn) and each of the other two telescopes has a MOS CCD X-ray camera in the 
focal plane$\sp{10}$ (EPIC-MOS1 and EPIC-MOS2). Each of these two cameras receive 
$\sim$50\%\ of the X-ray 
light leaving the exit plane of the corresponding X-ray telescope whereas the other 
half of the X-ray beam is intercepted by 
an array of grazing incidence gratings, part of the reflection grating 
spectrometers $\sp{11, 12}$ (RGS1 and RGS2).
The X-ray payload is complemented by the optical/UV monitor telescope 
OM$\sp{13}$. 

First light was received by the X-ray telescope equipped with the 
EPIC-pn camera on 19 January 2000 when it was pointed towards the 
30 Doradus region in the Large Magellanic Cloud$\sp{14, 15}$ with the 
other instruments following within days. The initial phase of  
operations and scientific observations was devoted to commissioning and calibration 
of the system, subsystems and the scientific instruments. Thereafter 
an extended period of performance verification followed until July 2000. 
Some of the scientific results of this first half year of XMM-Newton 
observations were published in a special issue of 
Astronomy \&\ Astrophysics$\sp{16}$. From July 2000 on observations 
of the Guaranteed Time Observing Program interleaved with observations 
of the first Guest Observer Program (AO-1) were carried out. 
Until early July 2001 data sets of $\sim$400 observations have been 
delivered to the original proposers.

The present paper focuses on the in-orbit performance of the three X-ray 
telescopes. The derivation of the performance parameters like 
point spread function (PSF), effective collecting area and vignetting of just the 
telescopes is not an easy task, since the measurements always involve 
the focal plane instruments and/or the reflection grating assembly which have 
their specific performance, so that the determination of the mirror PSF is 
intimately linked to the pixel size, photon pile-up, pattern pile-up, 
read-out mode, photon and charged-particle background, etc. of the CCD 
detectors. Likewise is the effective area determination sensitive to 
the CCD quantum efficiency, charge-transfer inefficiency and read-out 
mode of the CCDs, etc. Since the wings of the mirror PSF spread out over several 
independent and  
individual CCDs, which have different characteristics, in particular 
different background levels, the PSF is not easily established from 
a few imaging observations, likewise is the vignetting. Calibration 
work is still in progress, but the core of the PSF is understood and the 
collecting area is being close to be understood. At low energies, i.e. 
$<$ 1 keV, the energy dependent transmisson of the various absorption 
filters used in front of the CCDs complicates the issue of 
calibration of the effective collecting area, which is not settled, yet. 
 
Independent of the still standing issues of calibration the observations 
which are at hand clearly demonstrate that the scientific instruments 
on XMM-Newton are fulfilling 
their scientific capabilities qualitatively and quantitatively to the level 
the 
design promised many years ago. 

After a short description of the telescopes' design and its criteria followed 
by an outline of the manufacturing process of the mirrors the in-orbit imaging 
performance of the three X-ray telescopes is presented and 
confronted with the full X-ray beam calibration 
measurements taken at the PANTER. The second half of the paper is used 
to illustrate the scientific capabilities of XMM-Newton summarizing 
a few observations and the scientific progress made by them. 

\section{THE X-RAY Telescopes}

\subsection{Design}

The design of the optics was driven by the requirement of obtaining the 
highest possible effective collecting area over a wide band of energies up 
to 10 keV and more with particular emphasis on the region around 
7 keV, in which the K lines of the astrophysically important iron appear. 
Thus, the mirror system has to utilize a very shallow grazing 
angle of $\sim$0.5$\sp{\circ}$ in order to provide sufficient reflectivity 
at high energies. 
%
In grazing incidence optics the effective area is increased by 
nesting a number of mirrors and thus filling the front aperture as far 
as possible. The 
nesting efficiency is determined by the mirror shell thickness and, 
in case of very low grazing angles for high energy optics, by the 
minimum radial mirror separation which is required for integration 
and alignment. 
%
The thinner the mirror shells are and the narrower the shells are spaced the 
larger is the collecting area. In the past very thin aluminum foils shaped to  
double cones have 
been used.
Thin-walled mirrors have also been produced by replication techniques 
and electroforming. Electroformed mirrors have been built up mostly in 
solid nickel, which, however, suffer from high mass.
For the design calculations the shell thickness was assumed to increase 
approximately linear with shell diameter to guarantee sufficient stiffness. 
A minimum radial shell separation of 1 mm was considered to be feasible 
to handle the integration of the shells in a package. Both the paraboloid 
section and 
the hyperboloid section were replicated as a single piece from a 
single mandrel, which limited the axial length of the total mirror to 60 cm. 
The length  
is shared evenly by the paraboloid and the hyperboloid. Although 
 longer mirrors would have provided a larger collecting area they were 
discarded because they  
were suspected to be very difficult to be removed from the mandrel. 
The production of paraboloid and hyperboloid in one piece avoids the 
problems created by the additional tasks of aligning and mounting 
if paraboloid and hyperboloid are separate. Given the length constraints 
imposed by the launcher the focal length was fixed at 7.5 m. The optimum 
design, which fulfills the collecting area constraints, was found by 
ray tracing. Gold was considered as baseline for the mirror surface 
coating; attempts to produce replicated mirrors with an iridum coating, 
which would have provided somewhat better high energy response, failed 
because the adhesive forces of iridium on the mandrel surface turned 
out to be too strong so that patches of the iridum layer tended to 
stick on the mandrel. Eventually the mirror wall thickness was fixed 
to 1 mm for the outermost shell with a diameter 
of 700 mm and at 0.5 mm for the innermost shell of 306 mm diameter. 
Each mirror assembly contains 58 shells; adding more shells is  
rather inefficient in building up more collecting area because of the mass 
penalty involved and the low gain in effective area. 

\subsection{Manufacturing} 

The production of nickel shells involves three steps. As for any replication 
a master mandrel of negative shape is produced for each of the 58 mirror 
shells. The mandrel is of solid aluminum covered with a thin layer of 
Kanigen, which is ground and polished to the precise negative shape 
of the required mirror. The mandrel surface is measured by  
metrology means, including profilometers of various trace lengths and 
interferometers. Ray tracing of the metrology data reveals that the 
XMM mandrels have an angular resolution of $\le$ 5 arcsec HEW and 
 an rms surface microroughness of $\le$ 0.5 nm. 
The mandrels are coated with a gold layer $\sim$200 nm thick. 
Each mandrel is put in an electrolytical bath in which a nickel layer 
is being built upon the gold surface until the required thickness has 
been reached. By cooling the nickel shell  separates from the 
mandrel, whereby the gold surface sticks to the electroproduced 
nickel shell because of the higher adhesion forces.
After separation the nickel shells are exclusively supported by 16 strings 
threaded through 16 tiny holes which are equally spaced around the 
circumference of the mirror closely below the rim of the paraboloid.  
After inspection of the  surface and measurement of  a couple of meridional 
profiles each mirror is mounted to a spider like support structure by 
glueing the parabola mirror into small grooves in each of the 16 arms of 
 the spider. This is done with the mirror 
oriented vertically and under optical control. 
The spider support structure at the paraboloid front is the only support 
structure of the mirror shells. There is no counterpart at the 
aft of the hyperboloids; they are not constrained or supported mechanically 
at all. A close view of one mirror module sector out of the 16 sectors is shown in 
Figure 1.
Further details about the production of the XMM telescopes are given by 
de Chambure et al.$\sp{17}$.

\subsection{The X-ray Baffle}

From an X-ray optical point of view a telescope is to be complemented by an 
X-ray baffle, which 
eliminates X-ray straylight from sources outside the field of 
view. Rays from outside the field of view can reach the sensitive 
area of the focal plane detectors by single reflection from the 
rear end of the hyperbola, if the source is at an off-axis angle 
between 20 arcmin and 80 arcmin, in case of XMM-Newton. Rays reflected just once from any one 
the parabolas cannot leave the mirror assembly because of the close packing of the 
mirror shells. Less highly nested telescopes like 
those on the Einstein, ROSAT and Chandra observatories employ 
radial vanes with one set inside and one set in front of 
each mirror to eliminate singly reflected rays. A set of vanes 
inside of the XMM-Newton mirror modules was excluded because 
of the high mechanical complexity involved, but it was found that 
an external multi-stage baffle could completely eliminate 
 singly reflected rays. The concept foresees 
a set of  concentric cylinders$\sp{18}$, one 
for each mirror shell, placed 
in front of the mirror system. Each ring has an annular width slightly 
thinner than the corresponding mirror wall thickness and a cylindrical height 
characteristic for each mirror by which the singly reflected rays are 
blocked. 
 Because of the mechanical complexity together with the 
limited space available 
the design 
was simplified to two parallel annular plane  sieve plates of equal thickness. 
Each plate consists of 58 rings and 16 radial struts. The plates are mounted 
co-axial to and co-aligned 
with the front aperture cross section of the 58 mirror shells. 
The off-set  of the two sieve plates from the front of the mirror 
system is 385 mm and 439 mm, respectively$\sp{19}$. 
The efficiency of the sieve plate system has been ray traced 
 and demonstrated to reduce 
the straylight level by a factor of 5 to 10 depending on the 
position in the focal plane. Figure 2 shows the image of a region 
of the Large Magellanic Cloud which is centered close to  the pulsar 
PSR 0540-69.3 and an associated supernova remnant. In the lower right 
the typical pattern of singly reflected rays from a bright 
point source outside 
the field of view can be seen. 
The level of stray light flux received in the image confirms 
the efficiency of the X-ray baffle expected from ray-tracing 
calculations for the nominal configuration.

A complete overview of the telescope with all its subsystems 
including the mirror door, the entrance optical baffle, 
the electron deflector and additional support structure is given in 
Ref. 17.
    
\subsection{X-ray calibration on ground}

During Phase C/D of the XMM program five X-ray mirror modules have been built
three of which have been selected for flight, i.e. FM2, FM3 and FM4. FM1 
and FM5 have been kept as spares, numbering of the modules is in 
chronological order of production. 
X-ray testing and final X-ray calibration including measurement of the
PSF both on-axis and off-axis for a few energies 
between 0.28 keV and 8.05 keV and the effective area as well as the  vignetting 
function for a series of energies covering the range from 0.28 up to 
17 keV has been performed in MPE's PANTER test facility between 
April 1997 (FM1) and July 1998 (FM4). 
MPE's PANTER originally built for the development 
and calibration of the ROSAT X-ray telescope was modified and partially 
rebuilt in 1992/93  to cope with the size of the XMM telescopes 
and quality assurance requirements. In simple terms, 
the facility consists of a 130 m long vacuum pipe of 1 m diameter. 
At the one end a micro-focus characteristic X-ray line source is installed 
and the use of various anti-cathode materials provides the required 
coverage of the energy band. 
The X-ray source has a diameter of $\sim$1 mm. 
At the other 
end a huge cylindrical vacuum 
 tank 12 m long and 3.5 m in diameter houses the X-ray telescope to be tested. 
The typical distance between 
source and mirror module midplane, i.e. the paraboloid hyperboloid 
intersection, is $\sim$124 m. With a focal length of 7.5 m the image of 
the X-ray 
source is produced at a distance of 7.97 m measured from the mirror 
module midplane. Given these distances the angular size of the X-ray 
source is $\sim$1 arcsec, which means that structures of the mirror 
surface finer than this cannot be resolved.

Both full aperture measurements and sub-aperture 
measurements have been performed. The PSFs and the effective collecting 
area were measured with the entire 
aperture of the mirror module illuminated, implying a beam divergence 
between 8.5 arcmin and 19.5 arcmin full width between 
the innermost and outermost mirror shells. The divergence of the 
illuminating beam implies that the rays reflected off the frontal one third of 
the parabolas do not intersect with the hyperbolas. Therefore the 
surface of this section of the parabolas is not imaged and does not 
contribute to the PSF. Furthermore the corresponding 
reflection loci on parabola and hyperbola differ from those for 
on-axis parallel beam illumination. 
Images have been recorded with both  a position sensitive 
proportional counter (PSPC) which is a copy of the PSPCs flown on-board 
of ROSAT$\sp{20}$ and an X-ray CCD provided by the X-ray astronomy group 
of the University of Leicester in the UK. Whereas the PSPC is well suited to 
measure effective area and the wings of the telescope PSF 
the CCD with its significantly better spatial resolution has been used 
to assess the details of the PSF core out to angular 
distances of $\sim$125 arcsec limited by the linear size of the CCD. 
The PSPC has a larger field of view and the PSF can be traced 
out to angular distances of 984 arcsec. The comparison of the 
CCD and the PSPC encircled energy functions show that the CCD images 
out to 125 arcsec contain 97\% \ of the total flux at 1.5 keV and 
93\% \ at 8.05 keV, respectively.

Measurements of the effective collecting area were also done using smaller 
apertures to reduce the beam divergence. The open area of such an aperture is 
defined by the azimuthal 
width of one sector of the mirror module, i.e. 22.5$\sp{\circ}$, and the 
radial width corresponds to roughly one quarter of the radial extent of 
the mirror assembly defined by the radii of the outermost and innermost 
shells. In this way the full aperture is sampled by 64 sub-apertures.    
The sub-apertures were realized by two steel plates with appropriate 
cut-outs, which were placed roughly 0.5 m in front of the mirror module.
 By rotating the two plates against each other like a fortune wheel 
or 'Gl\"ucksrad'
any one of the 64 sub-apertures 
could be set up. The maximum beam divergence is $<$1.4 arcmin which is much 
closer to a parallel beam configuration than for the full aperture illumination. 
 
\subsection{Point spread function}

On January 19 2000 the X-ray telescope FM2 saw "First Light" of 
the X-ray sky$\sp{14,15}$ followed by FM3 and FM4 two days later. 
Two different X-ray cameras are used to record the X-ray images. 
In the focal plane of FM2 a novel type of X-ray CCD based on pn-technology 
is used (FM2/pn)$\sp{8,9}$ whereas FM3 and FM4 each are equipped with 
a more conventional CCD based on MOS-technology$\sp{10}$ 
(FM3/MOS1 and FM4/MOS2). Unlike FM2, FM3 and FM4 each carry 
a reflection grating spectrometer (RGS)$\sp{11,12}$ by which about 
half of the X-ray beam is diverted from MOS1 and MOS2.  
After "First Light" the imaging performance of the three telescopes was
of immediate interest and additional observations were made during the 
commissioning phase of XMM-Newton until March 8 2000. Sources studied 
include the sources of the "First Light Field" to the Large Magellanic Cloud 
and 
the point sources EXO 0748-67, LMC X-3, PSR 0540-69 and PKS 0558-504. 
To circumvent pile-up problems in the CCDs only observations taken in 
"small window" mode were used for the early assessment of the in-orbit PSF. 
In "small window" mode, however,  
the observational field is limited in size and the radial extent 
of the PSF which can be studied is limited accordingly. The PSFs 
taken on ground can therefore only be compared with the in-orbit 
results out to the angular distance imposed by the "small window" 
size, which is $\sim$30 arcsec in radius. The wings of the PSFs are not 
accessible with these early 
measurements. Likewise, the encircled energy functions,  
i.e. the radially integrated PSFs, have been normalized to the value 
measured on ground 
at the angular distance corresponding to the 
"small window" size. 

A major difference between the conditions on ground and in orbit 
is the type of detector used in the focal plane, in particular the  
pixel size. The CCD used on ground had a pixel size of 25 $\mu$m equivalent 
to 0.65 arcsec. The CCDs used in orbit have a pixel size of 40 $\mu$m for 
the MOSs and 150 $\mu$m for the pn, which relates to 1.1 arcsec and 
4.1 arcsec, respectively. In particular the pn-CCD pixel size appears too 
large to resolve the core of the PSF directly and simulations still have to be 
done to demonstrate the compatibility with the ground measurements. 

The results of this early analysis have already been reported$\sp{21}$ 
and only the results for FM4 coupled to EPIC-MOS2 are repeated here 
in Figures 3 \& 4. Ref. 21 summarizes the imaging performance of all 
three telescopes. The data show that the performance in orbit is 
basically the same as it was on ground for all three telescopes as far as the PSF 
out to $\sim$30 arcsec is concerned.  
Work is in progress to extend the analysis to image radii $>$30 arcsec, 
which is hard because of the difficulty to assess the background 
in the cameras properly and because the individual CCDs in both 
the pn and MOSs cameras show different performance making the 
 relative adjustment among the CCDs tedious and cumbersome.  
However, given the excellent agreement of the PSFs between  the ground and 
in-orbit measurements as far as the core of the PSF is concerned there 
is no major difference for the PSF wings between ground and in-orbit 
performance expected. I have therefore merged the PSF ground measurements 
at PANTER taken with the CCD and the PSPC to construct the joint PSF and 
associated encircled energy function (EEF) out to image radii of 
984 arcsec. The results for FM3 are shown in 
Figures 5 \& 6. I expect the PSF and the EEF in orbit very close to the 
results shown. Figures 5 \& 6 illustrate the fairly weak dependence 
of the PSF and the EEF for energies of 1.5 keV and 8 keV. 
I don't expect this to be different for FM2 or FM4 given the ground 
measurements.

Fig. 7 shows the 2-D PSF of 
FM3 for a point source in terms of 
iso-brightness contours. 
The contours illustrate the capability to resolve structures down to a 
few arcseconds. This capability is convincingly demonstrated by Figure 8, 
which shows the X-ray image of the Castor A/B binary system together with 
YY Gem or Castor C, another binary,	 taken with the EPIC-MOS1 camera. 
Castor A and B are known to be separated by 3.9 arcsec, and they are clearly 
resolved in the image. Even the flaring of either of the two stars and the 
associated temporal change of the X-ray flux can be monitored and attributed 
to one or the other star$\sp{22}$. 

\subsection{Effective collecting area}

The design driver for the XMM-Newton telescopes' geometry was to 
achieve maximal area at low energies (2 keV) without sacrificing area 
at high energies ($\sim$7 keV). The design promises a collecting area 
of the mirror assembly approaching 1900 cm$\sp 2$ for energies up to  
150 eV, $\sim$1500 cm$\sp 2$ at 2 keV, 900 cm$\sp 2$ at 7 keV and 
350 cm$\sp 2$ at 10 keV for each of the three telescopes on-axis.
At 15 keV the area drops to $\sim$12 cm$\sp 2$. 
The effective area for each of the telescopes was measured in MPE's 
PANTER X-ray test facility illuminating the full aperture with 
characteristic line radiation between 0.28 keV and 10 keV and using 
a copy of the ROSAT PSPC as focal plane detector.  
The diameter of the PSPC in conjunction with the 
XMM-Newton telescopes corresponds to a field of view of 984 arcsec radius, 
so that at higher X-ray energies some fraction of the effective area 
is likely to be missed because rays are scattered outside of the 
field of view due to mirror surface microroughness.    
The results are compared with predictions from ray-tracing calculations 
taking into account the nominal mirror shell geometry, the test facility 
geometry (beam divergence) and using 
the optical constants of gold originally published by 
Henke et al.$\sp{23}$. Up-to-date data, which have been used here, are found in ref. 24. 
Figure 9 shows that the maximally achievable area appears not to have been met, 
but instead a deficit of $\sim$15\%\ on average is apparent. At fixed energy 
the spread in area among the four telescopes is relatively 
low of 2 to 3\%\ at most. 
The deficit has been analysed by Gondoin et al.$\sp{7}$, drawing upon 
 additional dedicated reflectivity measurements carried out at PANTER. 
The main contributors are identified 
as mirror edge deformations, causing additional obstruction in the PANTER divergent beam, 
 and reflectivity losses. 
The sub-aperture tests at PANTER using the 'Gl\"ucksrad' configuration 
 with the much better collimated beams confirmed the additional shadowing by the 
extended edges, as the effective area turned out to be higher. 
The deficit in area is reduced at all energies and the remaining 
deficit of $\sim$7\%\ on average is attributed to intrinsic reflectivity 
losses which amount to 3 to 4\%\ relative per reflection (c.f. 
Figure 10). At higher energies ($>$3 keV) the deficit becomes slightly 
larger than the average because of increasing mirror surface scattering. 
These measurements are the basis for the in-orbit calibration files for the 
XMM-Newton X-ray telescope effective area. Figure 11 illustrates the difference 
between a perfect mirror module of perfect reflectance and zero scattering and 
the in-orbit performance described by the current calibration file. 
The difference follows basically the 'G\"ucksrad' measurements except 
for the energies which are close to the M-absorption and L-absorption edges of 
gold, at which the area recommended by the calibration file differs significantly 
from expectation. In particular the region of the M-edges around  2.3 keV is 
of concern. In-orbit spectra of negligible statistical errors and simple shape like 
power laws taken with the EPIC cameras 
show strong residuals when fitted which indicates a collecting area different from the calibration file. 
Further work is required, also to investigate  whether the energy levels and the 
optical constants given in the literature are actually correct. Work by Owen et al.$\sp{25}$ 
indicates some difference.  

The real test of the accuracy of the calibration data comes via the comparison and fits 
to the spectra of well-known sources keeping in mind that it is not only the 
mirror but eventually the entire instrument including the focal plane camera which 
is being tested. Willingale et al.$\sp{26}$ report a maximum difference between 
previous measurements of the Crab Nebula and the results obtained with the EPIC-MOSs
of 5\%\ as far as the normalization of the spectrum is concerned. The pure statistical 
error might be even less. This supports very much the current calibration of the telescopes 
for energies above $\sim$0.8 keV, where the Crab spectrum is bright. 
A cross-calibration between the EPIC MOS cameras and the EPIC-pn camera in this energy 
band, i.e. for energies  $>$1 keV, has been derived by Warwick et al.$\sp{27}$ who 
did a simultaneous fit of all three EPIC cameras to the spectra of the 
supernova remnant G21.5-0.9. Both the slope of the power law spectrum and the normalization 
agree within the statistical error, which is 2\%\ and 5\%\, respectively.
It appears that the broad band calibration of the mirror modules' effective area 
is without problems; the clarification of details around the gold absorption edges and the calibration 
at low energies ($<$1 keV) are still pending. 
        
The decrease of the effective area with increasing field angle is described 
by the vignetting function, which is the ratio of the off-axis area and 
on-axis area at fixed energy. The assessment of the vignetting function is important to describe the 
flux of off-axis point sources and extended sources like supernova remnants and clusters 
of galaxies. Since the EPIC field of view is covered by several independent 
CCDs both for the MOS cameras and the pn camera, any one of which has different 
properties, the vignetting function is difficult to be assessed quantitatively. 
Some attempts have been made by putting G21.5-0.9, a supernova remnant 
roughly 5 arcmin in diameter, at various field positions. The result in terms of the 
vignetting function at 10 arcmin is shown in Figure 12 for the MOS 
camera. The data are compared with the predictions made by the ESA built SCISIM tool, 
a S/W package which simulates images and spectra by ray-tracing. Globally speaking there 
is agreement except at energies above 7 keV. Analyses are under way to establish 
the vignetting function at other off-axis angles and for the other two cameras. 
  
\section{Some science results}

The power of XMM-Newton is to provide spectrally resolved images over a broad energy band of moderate 
angular and spectral resolution but high dynamical range and high 
signal-to-noise ratio given the large collecting area of the 
telescopes and the high quantum efficiency of the CCDs. Furthermore, high signal-to-noise 
and high resolution spectra are delivered from the RGSs at low energies. 
The compilation of 56 papers in the January 2001 special issue of Astronomy \&\ Astrophysics$\sp{16}$
 resulting predominantly from observations made during  
 commissioning, calibration and performance verification demonstrate that these 
objectives have been met as far as the performance of the instruments is concerned. 
In the following I add a few more results, some of them being in the 
process of being submitted to astro-journals.

\subsection{The explosion fragments of the Vela supernova remnant}
A couple of extended X-ray emission regions outside of the general boundary of the Vela supernova
remnant have been discovered$\sp{28}$, which were suggested to be composed mainly of stellar fragments
of the progenitor star. The spectrum of fragment A measured with ASCA$\sp{29}$ indicates a significant 
overabundance
of Si with respect to all other lighter elements
 which is consistent with the view that this fragment
originated deep in the interior of the star, but a quantitatively analysis was difficult to do.
Figure 13 shows the XMM-Newton EPIC-pn image of fragment A$\sp{30}$, of which the head, 
10 arcmin x 5 arcmin in size,
is the brightest part. Both the tail, which extends back to the SNR boundary, and the head show
pronounced structure of surface brightness. Figure 14 is a display of the XMM-Newton 
EPIC-pn spectrum of the
total fragment revealing the presence of strong Mg and Si lines.
A preliminary analysis indicates  that the Si/O abundance ratio in
the head region is about ten times solar, and that of Mg/O about 4 times solar.
This results confirms that fragment A is associated with the ejecta. It is noted that there is no
indication of S-K H- or He-like lines. The upper limit of the Si/S abundance ratio is just 
consistent
with current core collapse supernova models. The definite determination of this ratio could 
be a crucial test. Observation times of $>$100 ks with XMM-Newton would be needed. One of the
immediate questions is how such an explosion fragment can survive over a distance of $\sim$20 pc
for $\sim$10000 years. Preliminary hydrodynamical models show that this is possible but that the
survival time is very sensitive to the ratio of the initial fragment mass to the ambient matter 
density.

\subsection{The Tycho SNR}

Unlike the Vela SNR Tycho is just 429 years old and generally considered to be the
remnant of a type Ia supernova. Figure 15 shows the XMM-Newton EPIC-MOS broad-band 
image of the
almost circular remnant with an outer annulus of fairly uniform surface brightness and highly 
knotty
structure further inside. 
This appearance has brought foreward the idea that the outer annulus is
associated with the blast wave shock and that the knotty ring contains the reverse shock heated
ejecta. Previous X-ray spectra have shown the presence of many H- and He-like K emission lines up
 to
the Fe and Ni K-lines, including a blend of Fe-L lines. With XMM-Newton spatially resolved spectra
have been taken$\sp{31}$, which show a highly non-uniform distribution. The south-eastern rim, which
appears to lead the remnant's expansion there, is dominated by three knots. The spectra of each of the
knots can be fitted with the same temperature and the same ionization timescale but the abundances
of Si, S and Fe differ remarkably. The most northern knot shows the highest abundances
of S and Si with some Fe, whereas
the most southern knot basically contains no Fe demonstrated
by the absence of both the Fe-K and Fe-L lines. In comparison with the other two knots a factor of
10 less Fe seems to have been mixed into the Si layer at the explosion when the knots were formed.
The enhancement of Si and S indicates that the matter is from the deeper ejecta and the abundance
variations point to a mixing of the deeper Fe layer changing from place to place. In that sense these
knots resemble the fragments observed in the Vela SNR, except that the Tycho knots are much younger and
may reflect more directly the explosion physics. The angular separation of the knots is
 $\sim$15$\sp{\circ}$ and the linear extent of each knot is $\sim$0.3 pc, which is remarkably
close to the elongation of fragment A of the Vela SNR despite their different ages.

The high energy X-ray continuum is rather regular over the
remnant and it peaks just behind the shock front defined by the radio emission, so that it is tempting
to attribute the emission to the shocked ambient medium. But the detailed spectral 
analysis$\sp{32}$ shows
that this outer region appears to be overabundant in Si and S (c.f. Figure 15),
which is difficult to reconcile with
the outer shock. Have we seen the region of the outer shock at all,
 or is it still fainter than the regions observed so far?
Maybe imaging observations at even higher energies may reveal the outer shock region.
The XMM-Newton spectra further towards the center of the remnant show the presence of 
argon and calcium, in particular in the regions where the remnant is brightest (Figure 16). But the 
Fe-K line, which requires the highest ionization temperatures, shows up in a region 
which is just half-way between the center of the remnant and its rim. This is very much in contrast 
to what is observed in the south-eastern knots where the Fe-K line appears close to the rim 
in some of them. Consequently, the ionization conditions, the Fe spatial distribution and, most 
likely, also the Fe velocity distribution are highly asymmetrical throughout Tycho, which 
probably is related to the explosion details.     

\subsection{N132D in the Large Magellanic Cloud and the question of the origin of the high energy continuum
in SNRs}

N132D, a supernova remnant in the LMC, is estimated to be $\sim$3000 years old. The remnant, which resembles
the Cygnus Loop in its general surface brightness appearance with intriguing filamentary structure
and an apparent blow-out to the north-east, has been imaged by Chandra in great detail.
Figure 17 shows the first
high quality broad-band X-ray spectrum integrated over the remnant obtained with the XMM-Newton 
EPIC-pn camera$\sp{33}$.
Prominent emission lines of Si, S, Ar, Ca and Fe are clearly observed. In itself this spectrum is amazing
because it shows the capability of XMM-Newton: the spectrum covers a flux range of 4.5 orders of
magnitude in one exposure. Figure 18 shows the high-resolution spectrum obtained with the XMM-Newton
RGS spectrometer$\sp{34}$, through which 31 emission lines could be identified. For the first time
lines from H-like N ($\# 24$) and C ($\# 31$) are seen and they might provide a link with UV observations.
They also allow to better constrain the electron density which is dominated by the 
elements of low atomic charge. So far, a direct measurement was possible only down to oxygen.  
Modeling of the RGS spectrum indicates a temperature of $\sim$0.6 keV with a definite upper limit
of 1 keV. With this temperature and the abundance of Fe apparent in the RGS spectrum in ionic states
from FeXVII to FeXXI essentially no
emission of  Fe-K is expected, contrary of what the EPIC-pn spectrum shows (c.f. Figure 17).
There must be a high temperature component which produces the high energy continuum and the
Fe-K line emission. Interestingly, emission from Fe ions of progressively higher ionic charges 
of up to FeXXIV
are not detected although within the energy range of RGS (strong lines expected at
10.6 and 11.2 \AA ), which means that
plasma of any intermediate temperature between 0.6 keV and a few keV is absent.
Furthermore, the spatial distributions of the low temperature and high temperature
components are totally different. Whereas the images taken in ion lines of low ionic charge 
show pronounced
filamentary structure, although different for instance in OVII and OVIII, the high energy
continuum and the Fe-K line distributions are fairly uniform across the remnant. There appears
to be a high temperature plasma filling the entire volume of the SNR to 
a large extent whereas the low temperature plasma in the filaments seems to be concentrated 
on or close to its surface.

The spatial distribution of the high energy continuum is not unique to N132D, a similarly uniform
distribution has also been found for Tycho and Cas A. It has been suggested that the high energy
continuum is at least
partly due to synchrotron radiation from highly relativistic electrons ($\sim$10's of TeV)
accelerated by diffusive shocks. The first example discussed in this context was SN 1006 and quite a number
of SNRs have been added to this category basically because of their power law type X-ray spectra.
Diffusive shock acceleration should dominate at the outer shock and the emission should be pronounced
there which is not observed for 
Cas A$\sp{35}$. If the high energy continuum is due to synchrotron radiation
the acceleration mechanism is more likely to be of turbulent nature acting throughout the remnant.

\subsection{SN 1987A}

The explosion of SN 1987A was one of the rare historical astronomical
events and it has triggered research in many astrophysical fields.
In soft X-rays it took a long time, actually almost 3.5 years,
to detect emission but since then it has been monitored
regularly by ROSAT$\sp{36,38}$ and more recently by Chandra$\sp{37}$ 
and XMM-Newton$\sp{38}$.
Figure 19 shows the light curve, which can be remarkably well fitted
with a t$\sp 2$ relation. Soft X-rays are expected from the
interaction of the supernova shock with matter left from the wind
of the blue supergiant stage of the progenitor inside the inner ring 
visible in the optical.
The t$\sp 2$ dependence seems to suggest some cylindrical
rather than spherical symmetry of the interaction, which would have
implications on the matter density of the wind.
The XMM-Newton data points tend to exceed
the t$\sp 2$ best fit. Whether this is an indication that the
shock wave has already reached and partly penetrated the inner
ring around SN 1987A remains to be seen. The Chandra images have
resolved the supernova$\sp{39}$ indicating that the shock wave has reached
the inner surface or is close to contact.
For converting count rate into flux values
the spectrum obtained with the XMM-Newton EPIC-pn camera (Figure 20)
has been used except for the Chandra data points.
The absence of Fe-L and Fe-K emission in the spectrum sets an upper limit
of the Fe abundance of 0.03 solar. The abundances of the lighter
elements are consistent with 0.5 solar except for Si and S, each of
which seems to be overabundant by a few times solar. If confirmed
these abundances will play an important role for the understanding
of the final evolutionary phase of the SN 1987A progenitor.

Due to the broad band coverage of XMM-Newton a tail in the spectrum has been 
discovered which adds another component to the spectrum. The tail spectrum is apparently 
non-thermal and can be best represented by a power law. Right now it is not 
obvious whether the power law, which has a photon index of -2.6, is the first sign of 
the putative central pulsar 
or whether it is associated with the remnant. But the non-thermal flux puts an upper 
of the pulsar luminosity of $\sim$10$\sp{34}$ erg/s if the circumstellar debris is already optically 
thin at X-ray energies or $>$10$\sp{36}$ erg/s if the intrinsic absorption column density 
is $>$10$\sp{23}$ cm$\sp{-2}$. Alternatively, if the power law spectrum is associated with the remnant 
we might see the development of X-ray synchrotron radiation at an early phase or the bremsstrahlung 
of non-thermal electrons. Further monitoring is required and deeper observations should  be taken, 
in particular to look for a different time evolution of the two spectral components.

\section{Conclusions}

The XMM-Newton Observatory is performing well. The X-ray telescopes are doing the job they were 
designed for; in fact the angular resolution is significantly better than originally required. 
The scientific instruments are delivering a huge amount of new data of outstanding quality 
and it is without any question that many new discoveries are already on the door step. 

\section*{REFERENCES}

\begin{enumerate}
\item "A Cosmic X-ray Spectroscopy Mission", Proc. of a Workshop 
      held in Lyngby, Denmark on 24-26 June 1985, {\it{Proc. ESA}} {\bf{SP-239}}, 1985.
\item B. Aschenbach, O. Citterio, J.M. Ellwood, P. Jensen, P. de Korte, A. Peacock and 
      R. Willingale, "The High-Throughput X-ray Spectroscopy Mission", Report of the 
      Telescope Working Group, {\it{Proc. ESA}} {\bf{SP-1084}}, 1987.
\item B. Aschenbach and Heinrich Br\"auninger, " Grazing Incidence Telescopes for ESA's 
      X-ray astronomy mission XMM", in {\it{X-ray Instrumentation in Astronomy II}}, 
      L. Golub, ed., {\it{Proc. SPIE}} {\bf{982}}, pp. 10-15, 1988.
\item B. Aschenbach, "Re-Design of the XMM Optics", {\it{Techn. Note XMM/O/MPE}}, 1987.
\item "XMM - Special Issue", ESA Bulletin 100, ESA Publications Division, Noordwijk, The Netherlands, 
      1999.
\item P. Gondoin, P., B. Aschenbach, M. Beijersbergen, R. Egger, F. Jansen {\it{et al.}},
      "Calibration of the XMM Flight Mirror Module, I - Image Quality",
      {\it{Proc. SPIE}} {\bf{3444}}, pp. 278 - 289, 1998.
\item P. Gondoin, P., B. Aschenbach, M. Beijersbergen, R. Egger, F. Jansen {\it{et al.}},
      "Calibration of the XMM Flight Mirror Module, II - Effective Area",
      {\it{Proc. SPIE}} {\bf{3444}}, pp. 290 - 301, 1998.
\item L. Str\"uder, N. Meidinger, E. Pfeffermann, R. Hartmann, H. Br\"auninger {\it{et al.}},
      "X-ray pn-CCDs on the XMM-Newton Observatory", 
       {\it{Proc. SPIE}} {\bf{4012}}, pp. 342 - 352, 2000.
\item L. Str\"uder, U. Briel, K. Dennerl, R. Hartmann, E. Kendziorra {\it{et al.}},
      {\it{Astron. \&\ Astroph.}} {\bf{365}}, L18 -L26, 2001.
\item M. J. L. Turner, A. Abbey, M. Arnaud, M. Balasini,M. Barbera {\it{et al.}},
      "The European Photon Imaging Camera on XMM-Newton: The MOS cameras", 
      {\it{Astron. \&\ Astroph.}} {\bf{365}}, L27 -L35, 2001.
\item J. W. den Herder, A. J. den Boggende, G. Branduardi-Raymont, A. C. Brinkman, J. Cottam {\it{et al.}},
      "Description and performance of the reflection spectrometer on board of XMM-Newton", 
      {\it{Proc. SPIE}} {\bf{4012}}, pp. 102 - 112, 2000.
\item J. W. den Herder, A. C. Brinkman, S. M. Kahn, G. Branduardi-Raymont, K. Thomson {\it{et al.}},
       {\it{Astron. \&\ Astroph.}} {\bf{365}}, L7 -L17, 2001.
\item K. O. Mason, A. Breeveld, R. Much, M. Carter, F. A. Cordova {\it{et al.}},
      "The XMM-Newton optical/UV monitor telescope",  
       {\it{Astron. \&\ Astroph.}} {\bf{365}}, L36 -L44, 2001.
\item U. G. Briel, B. Aschenbach, M. Balasini, H. Br\"auninger, W. Burkert {\it{et al.}},
      "In-Orbit Performance of the EPIC-PN CCD-Camera on Board XMM-Newton",
       {\it{Proc. SPIE}} {\bf{4012}}, pp. 154 - 164, 2000.
\item K. Dennerl, F. Haberl, B. Aschenbach, U. G. Briel, M. Balasini {\it{et al.}},
      {\it{Astron. \&\ Astroph.}} {\bf{365}}, L202 -L207, 2001.
\item "First Results from XMM-Newton", {\it{Astron. \&\ Astroph. Letters}} {\bf{365}}, 
      EDP Sciences, Les Ulis, France, 2001. 
\item D. de Chambure, R. Lain\`e, K. van Katwijk, and P. Kletzkine, 
      "XMM's X-Ray Telescopes", in "XMM - Special Issue", ESA Bulletin 100, ESA Publications Division, 
       Noordwijk, The Netherlands, pp. 30 - 42, 1999.
\item B. Aschenbach, "Design of an X-Ray Baffle-System", 
      {\it{Techn. Note XMM-TS-FMP004}}, 1996.
\item D. de Chambure, R. Lain\`e, K. van Katwijk, W. R\"uhe, D. Schink {\it{et al.}},
      "The X-ray Baffle of the XMM telescope: Development and Results", Intern. Symposium on the
      Optical Design and Production, Berlin 1999, {\it{EUROPTO Series 1999}}, paper no. 
      3737-53.
\item E. Pfeffermann, U.G. Briel, H. Hippmann, G. Kettenring, G. Metzner {\it{et al.}},
      "The focal plane instrumentation of the ROSAT telescope", 
      {\it{Proc. SPIE}} {\bf{733}}, pp. 519-532, 1986.  
\item B. Aschenbach, U. Briel, H. Br\"aunger,W. Burkert, A. Oppitz {\it{et al.}}, 
      "Imaging performance of the XMM-Newton X-ray telescopes", 
       {\it{Proc. SPIE}} {\bf{4012}}, pp. 731 - 739, 2000. 
\item M. G\"udel, M. Audard, H. Magee, E. Franciosini, N. Grosso {\it{et al.}}, 
      "The XMM-Newton view of stellar coronae: Coronal structure in the Castor X-ray triplet", 
       {\it{Astron. \&\ Astroph.}} {\bf{365}}, L344 -L352, 2001.
\item B. L. Henke, E. M Gullikson, J. C. Davis, Atomic Data and Nuclear Data Tables, Vol. 54,
      July 1993
\item http://www-cxro.lbl.gov/
\item A. Owens, S. C. Bayliss, P. J. Durham, S. J. Gurman, and G. W.  Fraser, 
      "Apparent discrepancy between measured and tabulated M absorption edge energies", 
      {\it{Astrophys. Journal}} {\bf{468}}, pp. 451 - 454, 1996.
\item R. Willingale, B. Aschenbach, R. G. Griffiths, S. Sembay, R. S. Warwick {\it{et al.}},
      "New light on the X-ray spectrum of the Crab Nebula", 
       {\it{Astron. \&\ Astroph.}} {\bf{365}}, L212 -L217, 2001.
\item R. S. Warwick, J-P. Bernard, F. Bocchino, A. Decourchelle, P. Ferrando  {\it{et al.}},
      "The entended X-ray halo of the Crab-like SNR G21.5-0.9",
       {\it{Astron. \&\ Astroph.}} {\bf{365}}, L248 -L253, 2001.
\item B. Aschenbach, R. Egger, and J. Tr\"umper, 
      "Discovery of Explosion Fragments outside the Vela Supernova Remnant Shock-Wave Boundary", 
      {\it{Nature}} {\bf{373}},  pp. 587 - 590, 1995.
\item H. Tsunemi, E. Miyata, and B. Aschenbach, 
      "Spectroscopic Study of the Vela-Shrapnel", 
      {\it{Publ. Astron. Soc. Japan}} {\bf{51}}, pp. 711 - 717, 1999
\item B. Aschenbach and E. Miyata, 
      "XMM-Newton observations of the Vela fragment A", 
      {\it{Astron. \&\ Astrophys.}} to be submitted, 2001.
\item A. Decourchelle, J. L. Sauvageot, M. Audard, B. Aschenbach, S. Sembay {\it{et al.}},
      "XMM-Newton observations of the Tycho supernova remnant", 
      {\it{Astron. \&\ Astroph.}} {\bf{365}}, L218 -L224, 2001.
\item T. F. X. Stadlbauer and B. Aschenbach, 
      "Spatially resolved spectroscopy of Tycho's SNR", 
      Proc. X-ray Symp. "New Century of X-ray Astronomy", Yokohama, Japan, March 2001, 
      to be publ. in {\it{Astron. Soc. Pac.}}, 2001.
\item K. Dennerl and A. Read, priv. communcication, 2000.
\item E. Behar, A. P. Rasmussen, R. G. Griffiths, K. Dennerl, M. Audard {\it{et al.}},
      "High resolution X-ray spectroscopy and imaging of the supernova remnant N132D", 
      {\it{Astron. \&\ Astroph.}} {\bf{365}}, L242 -L247, 2001.
\item J. A. M. Bleeker, R. Willingale, K. van der Heyden, K. Dennerl, J. S. Kaastra {\it{et al.}}, 
      "Cassiopeia A: On the origin of the hard X-ray continuum and the implication 
       of the observed OVIII Ly-$\alpha$/Ly-$\beta$ distribution", 
       {\it{Astron. \&\ Astroph.}} {\bf{365}}, L225 -L230, 2001.
\item G. Hasinger, B. Aschenbach, and J. Tr\"umper, 
      "The X-ray lightcurve of SN 1987 A", 
       {\it{Astron. \&\ Astroph.}} {\bf{312}}, L9 -L12, 1996.
\item D. N. Burrows, E. Michael, U. Hwang, R. McCray, R. A. Chevalier {\it{et al.}},
      "The X-ray remnant of SN 1987A", 
      {\it{Astrophys. Journal}} {\bf{543}}, L149 - L152, 2000. 
\item B. Aschenbach, 
      "Supernova remnants - the past, the present and the future", 
      Proc. X-ray Symp. "New Century of X-ray Astronomy", Yokohama, Japan, March 2001,
      to be publ. in {\it{Astron. Soc. Pac.}}, 2001.
\item D. N. Burrows, E. Michael, U. Hwang, R. Chevalier, G. P. Garmire {\it{et al.}},
      "Early Results from Chandra Observations of Supernova Remnants", 
      {\it{Proc. SPIE}} {\bf{4012}}, pp. 91 - 100, 2000. 
\end{enumerate}

\vfill
\section{Figure captions}

\noindent
Fig. 1: Rear view of one sector of the
XMM-Newton mirror assembly FM2 showing
the 58 shells. The struts are elements of the electron deflector.
\hfill\break\vskip 0.25cm
\noindent
Fig. 2: 
X-ray image of the supernova remnant and pulsar PSR 0540-69.3 (close to image center) taken with the
EPIC-pn camera. In the lower right corner sections of rings appear which are due to singly
reflected rays from a point source outside the field of view to the lower right.
\hfill\break\vskip 0.25cm\noindent
Fig. 3: Point spread function of mirror module FM4 (solid line: in orbit,
dotted/dashed  line: on ground). 
\hfill\break\vskip 0.25cm\noindent
Fig. 4: Encircled energy function 
of mirror module FM4 (solid line: in orbit,
dotted/dashed  line: on ground).
\hfill\break\vskip 0.25cm\noindent
Fig. 5: Point spread function PSF 
of mirror module FM3 measured at PANTER combining the CCD and PSPC
measurements (solid line: 1.5 keV,
dotted line: 8 keV); PSF is normalized to unity for the flux
within 984 arcsec image radius, both for 1.5 and 8 keV. The dip around 400 arcsec 
is an artifact introduced by the PSPC.
\hfill\break\vskip 0.25cm\noindent
Fig. 6: Encircled energy function EEF 
of mirror module FM3 measured at PANTER combining the CCD and PSPC
measurements (solid line: 1.5 keV,
dotted line: 8 keV); EEF is normalized to unity for the flux
within 984 arcsec image radius, both for 1.5 and 8 keV.
\hfill\break\vskip 0.25cm\noindent
Fig. 7: Surface brightness of a point source image taken with mirror
module FM3 in orbit. The size of the frame is 44" by 44".
Contours are stepped by a factor of two in brightness$\sp{21}$. 
\hfill\break\vskip 0.25cm\noindent
Fig. 8: X-ray image of the Castor A/B and YY Gem system taken with the FM3 EPIC-MOS1 instrument;
Castor A and B, which are separated by 3.9 arcsec are
resolved.$\sp{22}$
\hfill\break\vskip 0.25cm\noindent
Fig. 9: Apparent effective area deficit of the XMM-Newton mirror modules
FM1 (triangle), FM2 (+), FM3 (x) and FM4 (square) measured at PANTER
with the full aperture illuminated.
\hfill\break\vskip 0.25cm\noindent
Fig. 10: Effective area deficit of the XMM-mirror modules
FM1 (triangle), FM2 (+) and FM3 (x) deduced from 64 sub-aperture measurements
using the 'Gl\"ucksrad' configuration, which provides a close to 'infinite source distance'
configuration. Maximal area data assume perfect reflectance and nominal mirror module
geometry.
\hfill\break\vskip 0.25cm\noindent
Fig. 11: Difference between in-orbit effective area as in current
calibration file and perfect mirror module of an XMM-Newton telescope. Difference
is attributed to lower reflectance and larger scattering with
higher energies, except for the regions close to the M and L absorption edges of gold.
\hfill\break\vskip 0.25cm\noindent
Fig. 12: Measurement (data) and prediction by ESA's SCISIM tool (histogram)
for the vignetting function of an XMM-Newton X-ray telescope
at 10 arcmin of-axis; x-axis is in units of eV
(graph courtesy D. Lumb).
\hfill\break\vskip 0.25cm\noindent
Fig. 13:
XMM-Newton EPIC-pn raw image of the Vela fragment A.$\sp{30}$
\hfill\break\vskip 0.25cm\noindent
Fig. 14: The EPIC-pn spectrum of the Vela fragment A.$\sp{30}$
\hfill\break\vskip 0.25cm\noindent
Fig. 15: (bottom): XMM-Newton EPIC-MOS image of Tycho's
supernova remnant$\sp{31}$; radius of the remnant $\sim$240 arcsec; 
 (top): EPIC-pn spectrum of the rim (image radius r$>$ 228 arcsec) in the 
south-west.$\sp{32}$
\hfill\break\vskip 0.25cm\noindent
Fig. 16: Two EPIC-pn spectra from the south-west for image radii 
190 arcsec $<$ r $<$ 220 arcsec (top) and 135 arcsec $<$ r $<$ 155 arcsec 
(bottom).$\sp{32}$
\hfill\break\vskip 0.25cm\noindent
Fig. 17: XMM-Newton EPIC-pn spectrum of the supernova remnant
N132D$\sp{33}$.
\hfill\break\vskip 0.25cm\noindent
Fig. 18: The spectrum of N132D resolved with the RGS
spectrometer$\sp{34}$.
\hfill\break\vskip 0.25cm\noindent
Fig. 19: 0.5 -- 2.0 keV light curve of SN 1987A
compiled from data obtained with ROSAT (open squares)$\sp{36,38}$,
Chandra (triangles)$\sp{37}$ and XMM-Newton (filled squares)$\sp{38}$.
\hfill\break\vskip 0.25cm\noindent
Fig. 20: Broad band spectrum of SN 1987A obtained with the
EPIC-pn camera$\sp{38}$.

\end{document}